\documentclass[11pt]{article}

\usepackage[margin= 1 in]{geometry}

\usepackage{graphicx}
\graphicspath{ {./images/} }

\usepackage{lipsum}
\usepackage[dvicolors]{xcolor}
\usepackage{textcomp}

\usepackage[utf8]{inputenc}
\usepackage[english]{babel}

\usepackage[backend=bibtex, style= nature]{biblatex}
\usepackage{amsmath} 

\usepackage{caption}
\usepackage{textgreek}

\usepackage[switch]{lineno}

\DeclareUnicodeCharacter{2212}{-}

\begin{document}

\title{Selectively biased tri-terminal vertically-integrated memristor configuration}

\author{Vasileios Manouras$^*$ $^1$ , Spyros Stathopoulos$^1$, Alex Serb$^1$, and Themis Prodromakis$^1$}

\date{%
$^*$Corresponding author\\%
$^1$Centre for Electronics Frontiers, Electronics and Computer Science, University of Southampton, SO171BJ, Southampton, United Kingdom\\[2ex]}

\maketitle

\begin{abstract} 

Memristors, when utilized as electronic components in circuits, can offer opportunities for the implementation of novel reconfigurable electronics. While they have been used in large arrays, studies in ensembles of devices are comparatively limited. Here we propose a vertically stacked memristor configuration with a shared middle electrode. We study the compound resistive states presented by the combined in-series devices and we alter them either by controlling each device separately, or by altering the full configuration, which depends on selective usage of the middle floating electrode. The shared middle electrode enables a rare look into the combined system, which is not normally available in vertically stacked devices. In the course of this study it was found that separate switching of individual devices carries over its effects to the complete device (albeit non-linearly), enabling increased resistive state range, which leads to a larger number of distinguishable states (above SNR variance limits) and hence enhanced device memory. Additionally, by applying a switching stimulus to the external electrodes it is possible to switch both devices simultaneously, making the entire configuration a voltage divider with individual memristive components. Through usage of this type of configuration and by taking advantage of the voltage division, it is possible to surge-protect fragile devices, while it was also found that simultaneous reset of stacked devices is possible, significantly reducing the required reset time in larger arrays.  
    
\end{abstract}

\section{Introduction}

Since their conception \cite{Chua} and eventual fabrication \cite{Strukov} memristive devices have been extensively studied in an effort to incorporate them in circuit designs \cite{Papandroulidakis} or in novel neuromoprhic computing setups \cite{Yoon}. Increased integration density, which resulted from incorporation in circuits, uncovered issues which impede nominal device operation, such as sneak path currents. A popular mitigation strategy has been the usage of new topologies, such as complementary resistive switching (CRS) \cite{Lee}, 1 selector 1 memristor (1S1M) \cite{Woo}, 1 diode 1 memristor (1D1M) \cite{Liu} configurations. Other theoretical propositions include using a complementary memristor array composed by two anti-serially connected memristors \cite{Jung}. At the same time, these technologies also protect from unwanted switching or destructive breakdown of fragile devices, by enforcing a (soft) compliance current through the multiple layers used to fabricate them. 

On a separate front, efforts are being directed towards increased range in tuneability of devices \cite{Stathopoulos_multibit}, especially in settings of neuromorphic computing, were synaptic weight management is of vital importance. By using memristors as standalone components, analogue reconfigurable circuits \cite{Serb_reconfig} and threshold logic gates \cite{Papandroulidakis} can be fabricated, among other possible circuitry. One commonly referenced `device ensemble' is the memristive fuse \cite{Isah, Gelencser}, which elaborates on the expected behavior of two memristors in series. Specifically, it has been previously pointed out \cite{Serb_fuse} that memristive fuse is the natural extension of complementary switching when states in both devices exhibit analogue behaviour. In all of these `hardwired ensemble' cases it is not possible to gain insight into the workings of the separate layers of the ensemble, and fine tuning of states is a critical issue which has not yet been resolved. 

\begin{figure*}
    \centering
    \includegraphics[scale = 0.95]{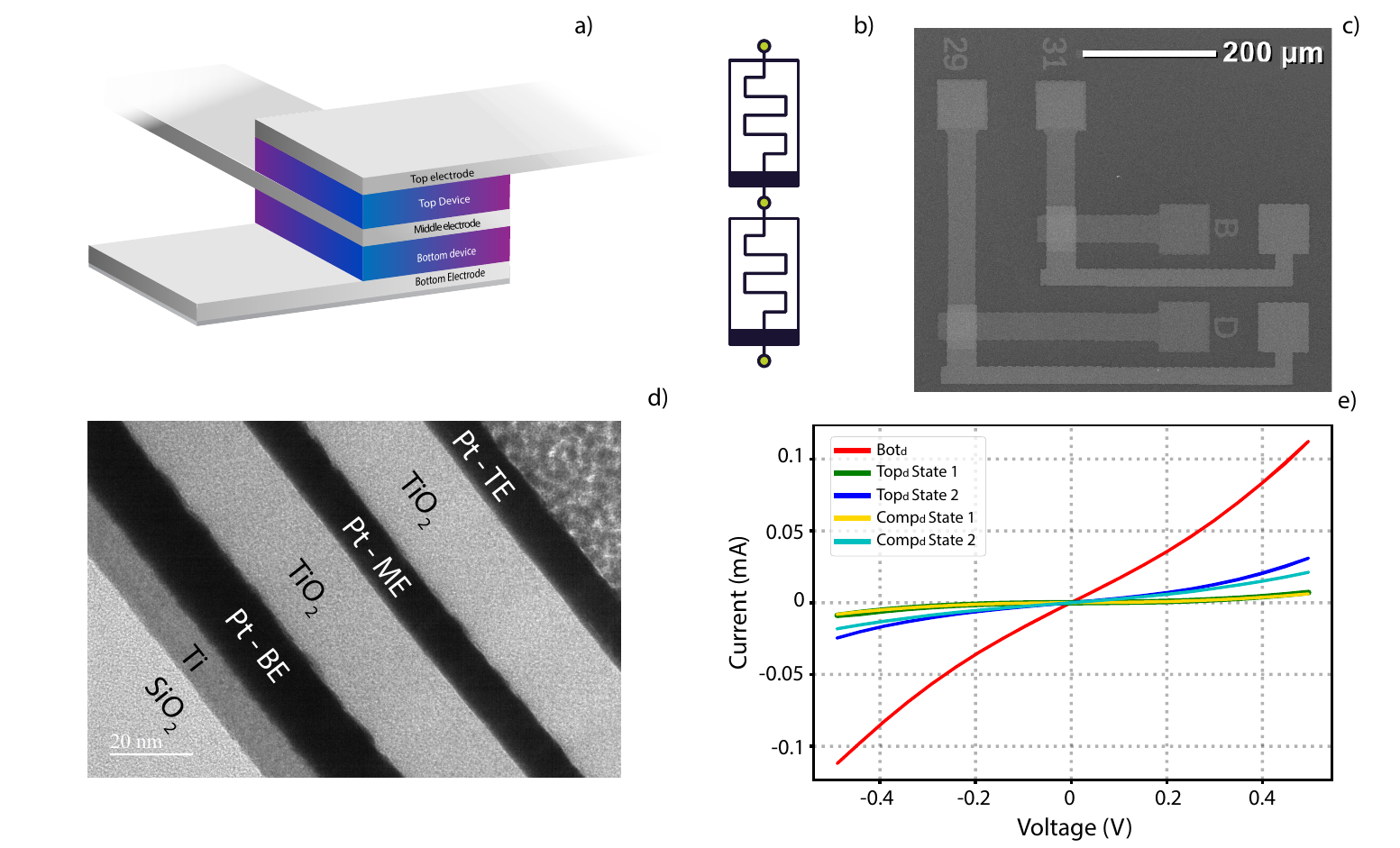}
    \caption{Physical and electrical characteristics of MIMIM devices. (a) graphical depiction of device topology, while (b) electrical equivalent model ('serial' connection). (c) SEM image of the fabricated devices, illustrating the 3 electrodes and their overlap areas which constitute the devices. (d) TEM image of the overlap area, showing the vertical stacking of the fabricated devices. (e) non-switching IVs of a stack, whereby the IV of the total stack is dependent on the IVs of the top and bottom devices, and the resulting voltage division between the two. In this case, two distinct states for the top device were registered and the switching impact on the total stack has been recorded.}
    \label{fig 1}
\end{figure*}

Here, we propose a three terminal device, 1M1M topology, which consists of two memristors fabricated vertically, with an interleaved, but accessible common middle electrode. The case for three terminal connected devices has already been theoretically made where this configuration is used to propose neuromoprhic or conventional circuits \cite{donglian, Jung}. This configuration allows probing of the whole stack, while the ability to tune both memristors separately allows fine tuning of the compound stack's resistive state. Other expectations from this type of configuration include increased tuneability, simultaneous switching of both devices and an inherent surge protection of devices due to the stack functioning as a voltage divider, improving resilience. With regards to fine tuneability, `super-resolution' memristive ensembles represent an idea that has been explored theoretically, with cells consisting of two memristors connected in parallel \cite{James}, while here we are implementing it in a serial connection topology. This configuration allows both higher resolution and higher memory density per unit area. Both are achieved via manipulation of individual devices with different switching ranges, enabling the set-up of a coarsely-set resistive state originating from one device plus fine resistive tuning around that state by utilizing the second device. Notably, even for devices with similar resistive ranges and tuning tolerances, independent control of resistive states of two devices can achieve better control of the aggregate resistive state. Finally, the superposition of resistive states across both devices, broadens the overall tuning range of the aggregate state, providing further memory density (more available states). For simplicity we shall refer to these effects simply as `super-resolution' in the rest of the paper. Control of forming \cite{Yang} and/or device manufacture can lead to both 'serially' and `anti-serially' connected devices which can exhibit distinct computational behaviours. The difference being that in a `serial' stack, application of voltage of some chosen polarity will elicit the same response from both devices in the stack (both will either increase or decrease resistance).

\section{Sample preparation}

Final device structure can be summarized as a metal-insulator-metal-insulator-metal (M-I-M-I-M). The two fabricated prototypes were: Pt(BE)/TiO$_x$(AL1)/Pt(ME)/TiO$_x$(AL2)/Pt(TE) and Pt(BE)/[TiO$_x$/Al$_2$O$_3$](AL1)/Pt(ME)/TiO$_x$(AL2)/Pt(TE). Abbreviations in the parentheses denote the function of each layer within the device stack. A complete SEM image of all three electrodes, their access pads and the overlap area can be seen in Fig \ref{fig 1}(c), while a visual representation of the cross sectional area is available in Fig \ref{fig 1}(a). Finally, Fig. \ref{fig 1}(b) is a depiction of the electronic circuit which this configuration offers, with each of the three access points depicted as green circles. A TEM image of the cross-sectional area (Fig. \ref{fig 1}(d)) depicts the internal composition of the devices ensuring film uniformity and quality.

Memristive devices are initially fabricated as M-I-M capacitors, and they transition into memristors after an electroforming step. This electorforming step is carried out independently for each memristive device in the stack, by making use of the ME. Electroforming and measurements are carried out by using the ArC One platform , previously developed for memristor characterization \cite{Berdan}. This enables the use of highly controllable pulsing schemes for gentle and controllable electroforming, as well as controllable and gradual state switching, by utilizing pulse trains instead of IVs. All testing in this work is current-compliance free. The electroforming procedure comprises of 4 repeated steps to ensure successful device activation and has been covered extensively as part of a larger testing procedure in a previous publication \cite{Manouras}. Initially an IV sweep is carried out to ensure good connection to the device and to observe typical rectifying behavior of unformed devices. Afterwards, the electroforming step takes place, in which a pulse train of steadily increasing voltage is sent to the device, until a sharp drop in resistance (to below a specified threshold) triggers the stop condition of the pulsing algorithm (Supplementary Figure S1). A new IV curve of the device is then obtained, which must be pinched, and resistive switching must be observed, for it to be considered as formed. Finally a retention step is carried out to ensure nonvolatile behavior.

\begin{figure*}
    \centering
    \includegraphics[scale = 0.9]{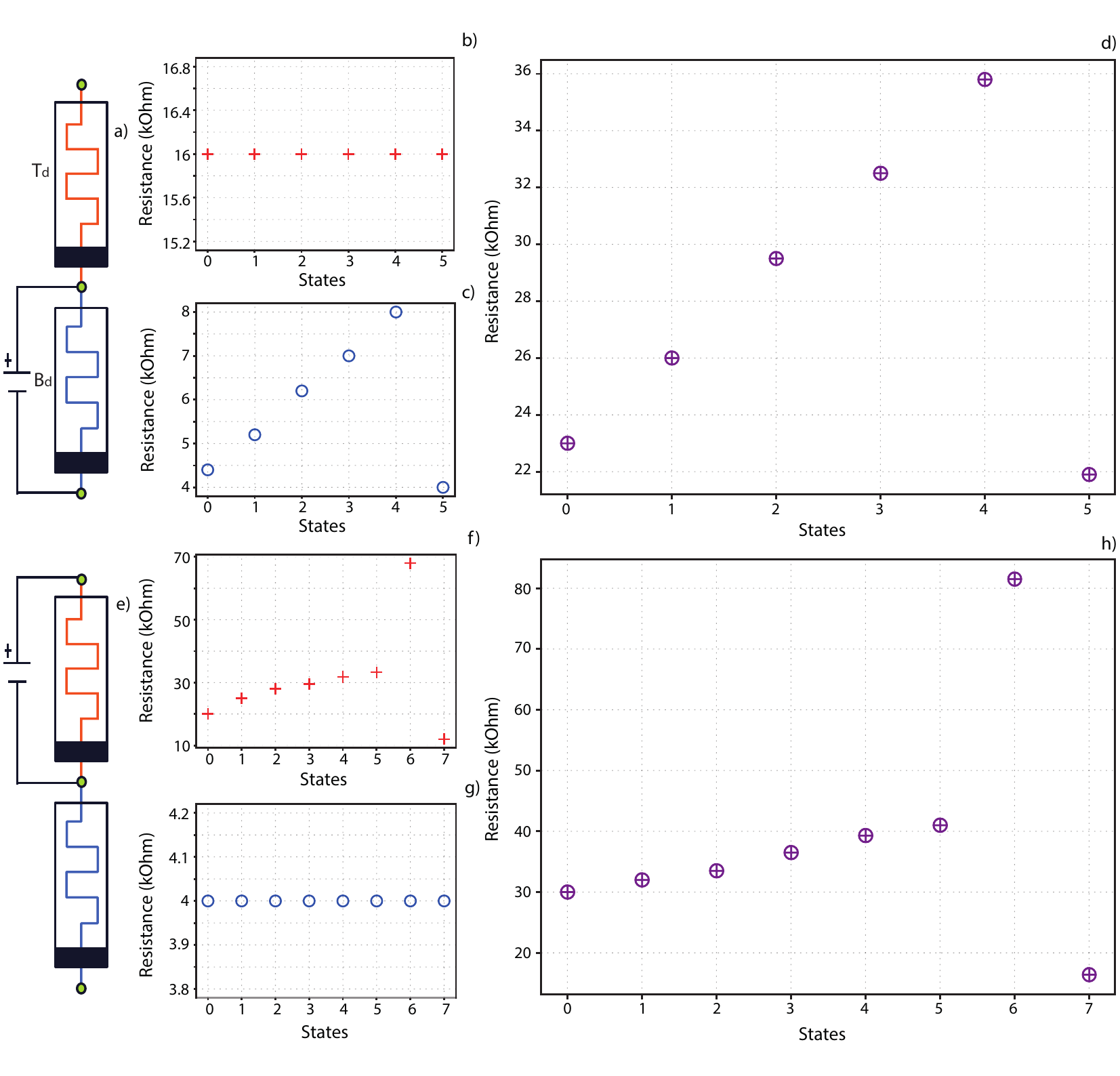}
    \caption{Total stack switching when the bottom device is switched (a-d) and when the top device is switched (e-h). Subfigures (a) and (e) depict the equivalent circuit during programming, where the switching and non-switching components can be distinguished. Subfigures b), c), f) and g) show the corresponding states of the devices in each step of the experiment, with (b) and (c) corresponding to the Top and Bottom devices of subfigure (a), while (f) and (g) correspond to the respective devices in subfigure (e). Evidence from (b) and (g) shows that the secondary device remains stable, while the switching device is being programmed in (c) and (f) respectively. Subfigures (d) and (h) result from measuring the resistive state of the total stacks ((a) and (e)). In both cases it is evident that total stack resistive states follow the curve set by the switching device, but the relation is not simply additive (in both cases a change of $X \Omega$ in the switching device leads to a change $Y>X \Omega$ at the output).}
    \label{fig 2}
\end{figure*}

The forming procedure was applied \emph{individually} for both top and bottom devices. After forming, the possible measurement combinations available are Top device (Td), Bottom device (Bd) and the Complete device (Cd) or `Total stack'. Measurements of each configuration were carried out by positioning the probing needles on the TE and ME, ME and BE, or TE and BE electrodes respectively. The third, unused, electrode was not engaged (floating terminal), to avoid any impact on measurement. Fig \ref{fig 1}(e) displays non-switching IVs of a fully formed device, with 2 states used on the Top device, impacting the complete device IV, respectively.

Here, two types of measurements were carried out. The first type assesses switching impact on the Total stack due to only one device switching. In this case, a resistive readout was taken of the non-switching component and of the total stack, while a pulse train was used to induce switching in the switching component. This was followed by a subsequent resistive readout. For the second type of measurements, concerning the switching of the Total stack, a pulse train was applied across the Top and Bottom electrodes, which could lead to switching in both layers or only one of them. Due to the inherent non-linearity of memristive device IVs and to ensure consistency and comparability between measurements, whenever a resistive state value is quoted, this refers to static resistance under 0.5V read-out voltage unless otherwise specified.

\section{Characterization}

Device characterization has been initially carried out by switching one of the two memristors and observing the effect on the Total Stack and then switching the Total stack as a whole and observing the effect this had on the two devices it was composed of. As expected, full-stack resistance changes when either of the devices switches. This can be seen from the non-switching IVs recorded in Fig \ref{fig 1}(e), where a state change in the top device led to a corresponding state change in the Total stack. However, the total stack operates as a voltage divider and due to: a) the redistribution of voltage within the divider due to generally unequal switching between constituent devices and b) the inherent non-linearity in device IVs, the final resistance of the total stack is dependent on, but not equal to, the sum of resistances of each individual device.

\subsection{Switching in single layer}

Individual switching was carried out independently on both of the devices comprising a total stack, as shown in Fig. \ref{fig 2}. Figure \ref{fig 2}(a,e) indicates which part of the device is switching in each set of measurements. In each case the non-switching device was completely unaffected, as seen in Figures \ref{fig 2}(b,f), confirming our ability to control devices independently. On the other hand, state switching of a device is transmitted to the full stack, as seen in Figures \ref{fig 2}(c,g), when compared with figures \ref{fig 2}(d,h) respectively. As a result of this behavior we can achieve high tuneability range of resistive states by finely tuning each one of the two separate devices, which will result in a combined range and potentially increased density/resolution of controllably achievable states in the final stack, increasing state resolution.

\begin{figure*}
    \centering
    \includegraphics{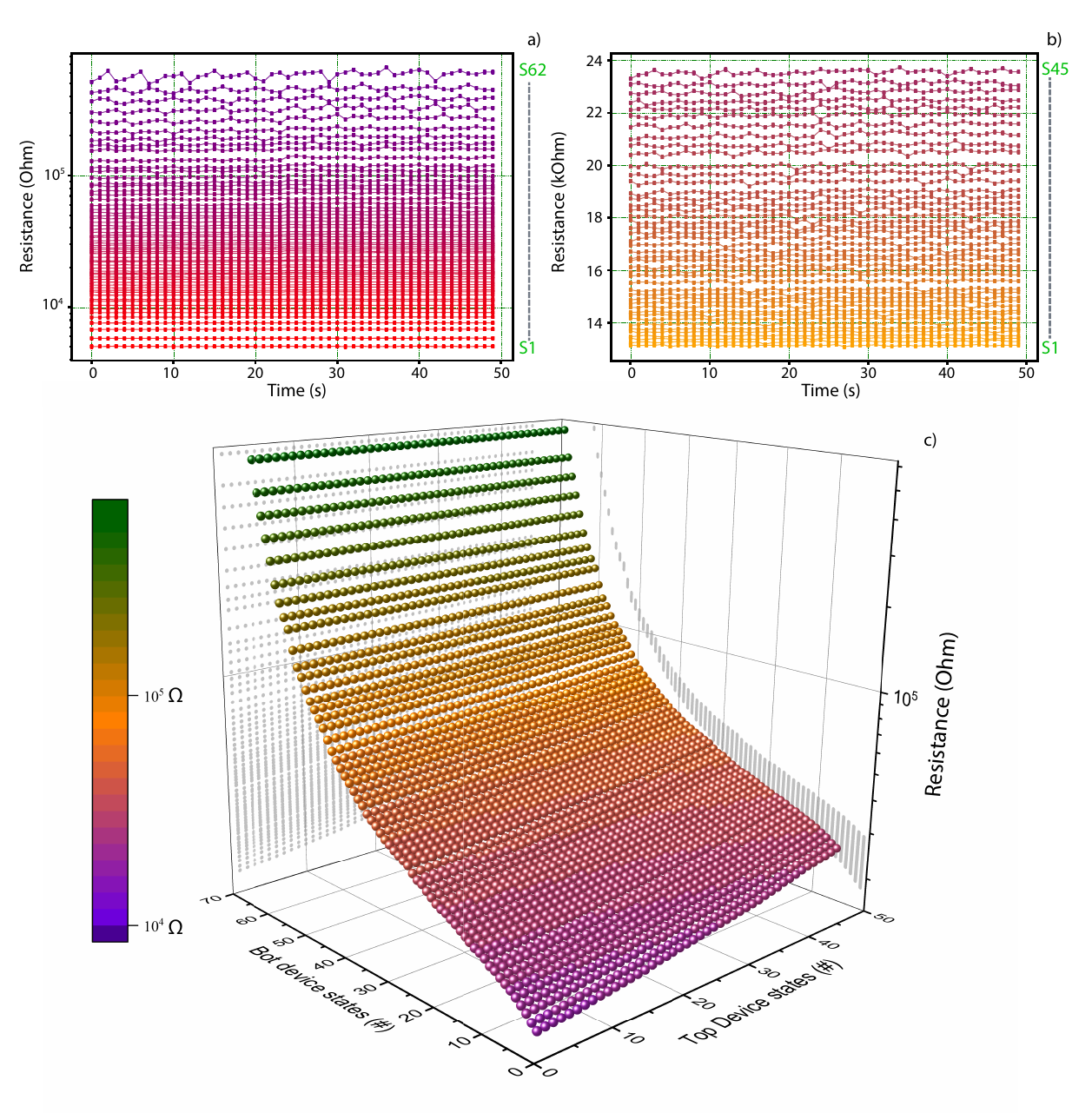}
    \caption{Improving memory density (number of memory states per device) and resolution per unit area of memristors by exploiting different resistive switching ranges in the top and bottom devices. (a,b) Compiled retention measurements of distinct resistive states in the Bottom (a) and Top (b) devices in the same Total stack, with data obtained by using a measurement routine outlined in previous work \cite{Stathopoulos_multibit}. The bottom device has a range of states which spans from single-digit kΩ to hundreds of kΩ. The top device ranges between approximately 10-30 kΩ. (c) All possible states that \emph{could} be achieved by finely tuning each separate device, calculated by using the sum of possible states of constituent devices. Resistive values shown here will inevitably deviate a small amount from the actual values due to voltage division, but the core concept of increased memory density remains unaffected. Multiple states may overlap (i.e. multiple combinations of top and bottom device resistive states lead to the same total-stack state), but total number of supported states will increase. This is indicative of the devices acting collaboratively in a coarse-fine tuning partnership; As seen in the projection on the right side of the 3D image, the top device would act as a fine tuner for the coarser bottom device states.}
    \label{fig 3}
\end{figure*}

A convenient way of exploiting this behaviour would be setting the aggregate stack state by using a device with large resistive state range and inter-state gaps for coarse tuning and a corresponding small range/gap device for fine tuning. To illustrate this, we utilize a multistate seeking protocol, which has been described at length in previous publications \cite{Stathopoulos_multibit, Stathopoulos_methodology}. In brief, the protocol returns the number of statistically distinguishable resistive states that a target device can achieve. The protocol is applied separately to both the top and bottom devices in the same stack, returning the states illustrated in Figures \ref{fig 3}(a) and \ref{fig 3}(b). The former depicts short retention runs for all stable states (62) that have been found in the Bottom device, spanning a 2-order-of-magnitude (500kΩ) resistive state window, while the latter shows stable states found in the top device (45), spanning a narrower resistive range (10 kΩ). For this experiment we consider two adjacent resistive states to be distinguishable if they are separated by 1$\sigma$ deviation. 

\begin{figure*}
    \centering
    \includegraphics[scale = 0.95]{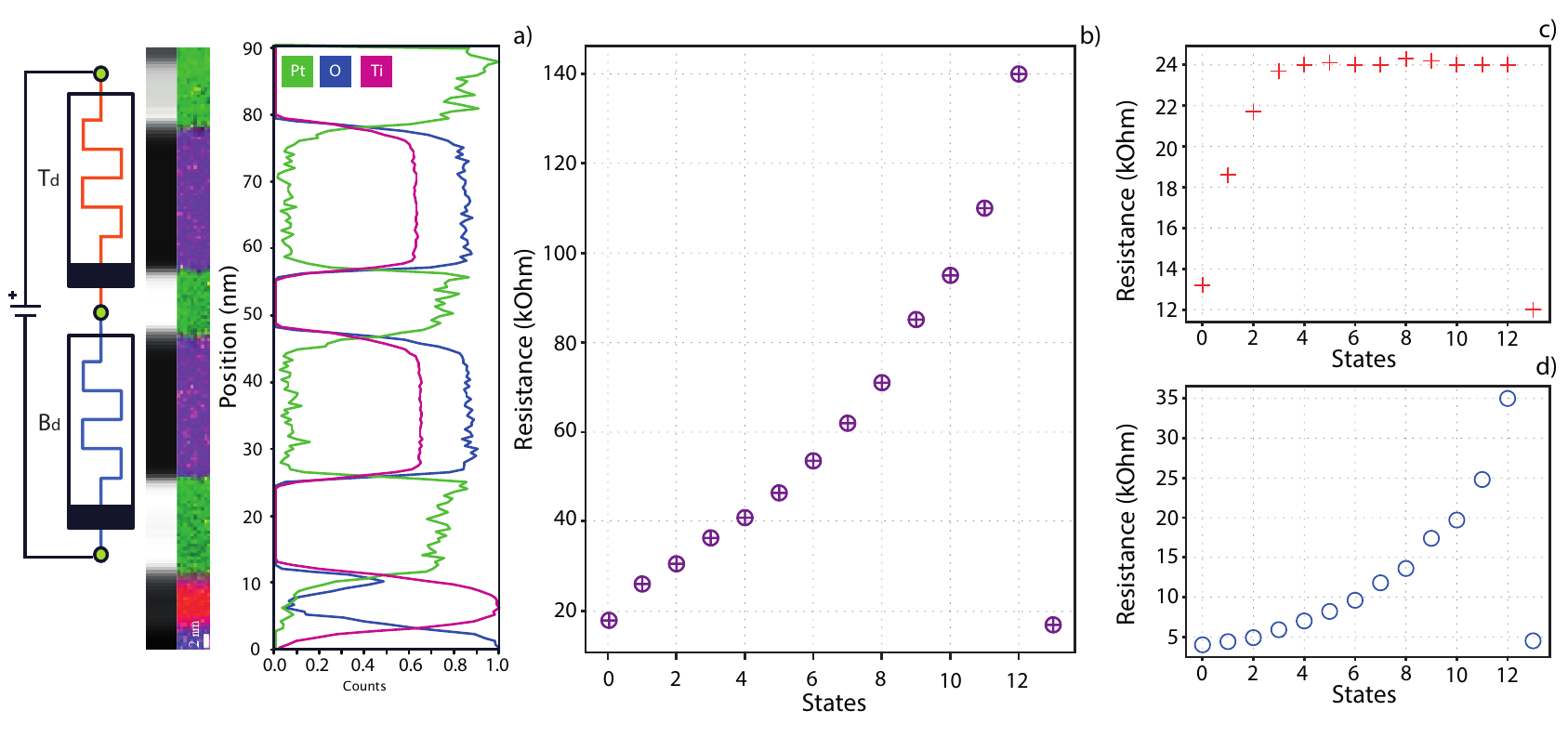}
    \caption{Simultaneous switching of memristors with the same switching polarity. Applied voltage is expected to have the same result on both devices, i.e, simultaneous resistive increase or decrese in both devices. (a) Schematic and an EELS image of the stack can be seen. (b) Total stack state evolution, where the first half of the change is driven mainly by the top device while the second half is driven mostly by the bottom device. (c,d) State evolutions for both devices, common x-axis. After the 4th state change, the top device saturates but the bottom device keeps increasing its resistance.} 
    \label{fig 4}
\end{figure*}

The 3-D map in Figure \ref{fig 3}(c) illustrates all states that could be obtained in the full stack, resulting from the sum of states in the two separately tested devices, which can be achieved by separately controlling each individual device. Due to inherent non-linearity and voltage division of devices, the actual resistive values would ,understandably, deviate from the calculated values, but the number of states would remain unaltered, as would the expected stack behaviour, due to selection of devices with different resistive ranges. This illustrates the point of a finely controlled super-resolution memristor. In this case, the individual devices have their states spread out in different patterns. The bottom device has a large number of states spread out over a range of 500 kΩ, while the top device has all its states within the boundaries of 10 kΩ. This in effect allows for the usage of the top device as a fine tuning mechanism, as can be seen in the rightmost 2-D projection of the 3-D map in Figure \ref{fig 3}(c). 

While many of the states presented here are duplicates, it is obvious that the combination of these two devices results in a state resolution which could not be achieved individually by any of the two devices.

\subsection {Simultaneous switching}

This section covers simultaneous switching of devices by using the top and bottom electrodes, TE and BE. This in effect enables the entire stack to work as a single device, but with the added functionality of being able to probe the individual components to ascertain where the switching took place and how it affected them. 

Two cases are studied in this section, the first is simultaneous switching of serially connected memristors, i.e. with the same switching polarity, while the second case is switching of antiserially connected memristors, i.e, alternate polarity.

Figure \ref{fig 4} depicts the first case , where both devices are switched in the same way. Subfigure a) depicts a schematic of the stack and measurement setup, along with an EELS image of the type of device used, which indicates the inner structure of the device from a materials point of view. Switching devices of comparable resistance simultaneously requires sufficient voltage reaching each device. This depends on the allocation of voltage in the potential divider that they form and the relation between the switching thresholds of the constituent devices as explained in \cite{Serb_fuse} The resistive readouts of the total stack are depicted in Fig \ref{fig 4}(b).

In the first 4 states, as seen in Figures \ref{fig 4}(c) and \ref{fig 4}(d), the top device exhibits a bigger resistive switch, presumably due to receiving a substantial majority of the divider voltage as indicated by the difference in nominal resistances between top device and Bottom device. After the "state 4" mark, the top device reached a threshold where the voltage applied was no longer enough to switch it to a higher state. This in turn left the bottom device able to continue its gradual switching towards higher resistive values, which naturally also led to a higher voltage share being directed to it. This resulted in the exponential increase in resistive value for the bottom device. Had the top device not hit a threshold it can be inferred that the switching of the bottom device would have been severely restricted.

Reading or switching devices by using the TE and BE of the stack has the added effect of doubling as a surge protection mechanism, as the voltage division means that devices will not receive more voltage than they can process, reducing the probability of device failure. 

The last part of the test saw the simultaneous reset of both devices by only sending one pulse through the device. As devices are very sensitive to voltage, especially when applying reset pulses to them, this approach has multiple benefits. First, due to voltage division-induced negative feedback the reset pulse is less likely to exceed the absolute maximum device limits, mitigating the risk of failure. Second, reset time is effectively cut in half. Indeed even if these stacks where to be used just as two conjoined, but individually operating memory cells, the simultaneous reset could still prove extremely useful, due to halving memory reset time. 

\begin{figure*}
    \centering
    \includegraphics[scale = 0.95]{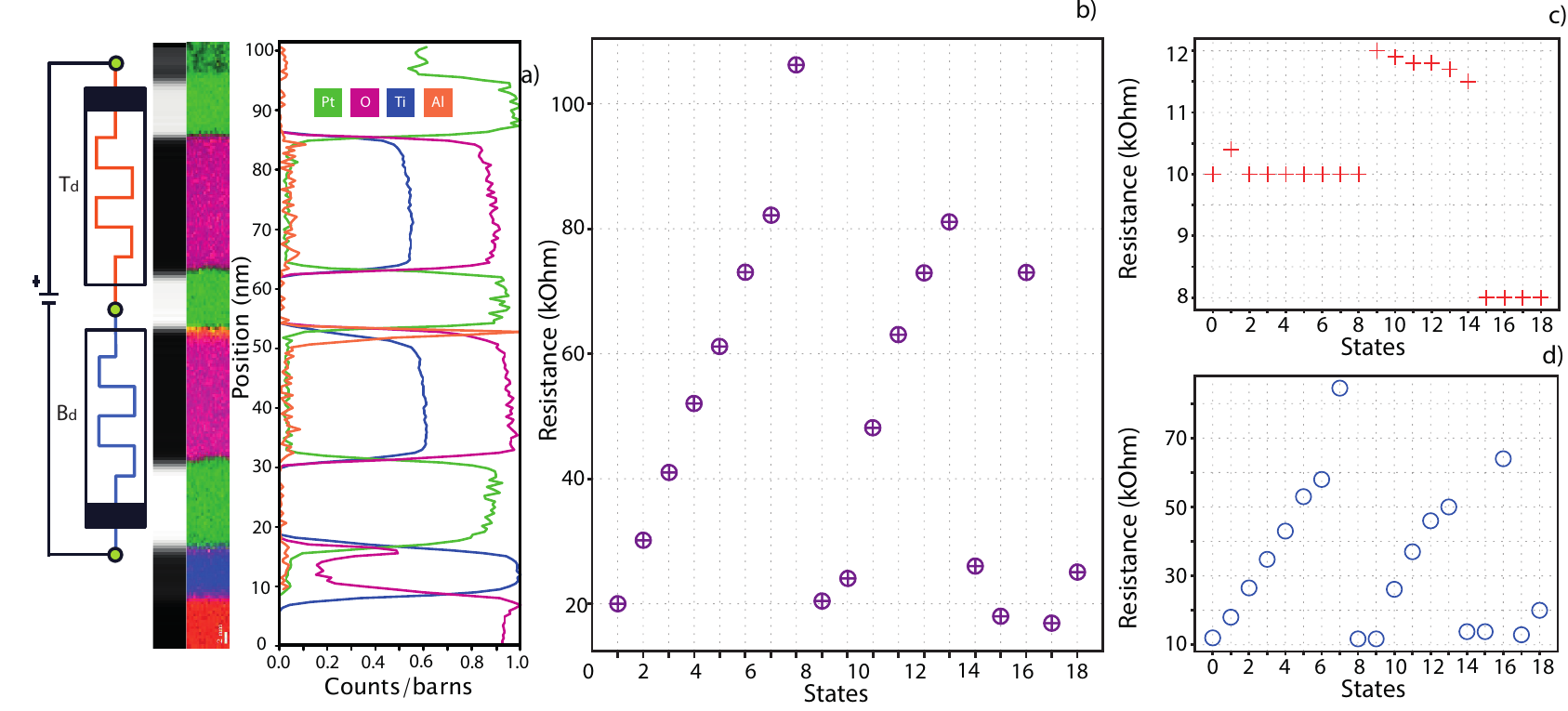}
    \caption{Simultaneous switching of antiserially connected memristors. In this case the voltage applied should have opposite effects on the devices, i.e. it could increase resistance in one and decrease it in the other. (a) Antiserial connection schematic and EELS image of the stack. In this case the bottom device used was a bilayer stack to ensure cross-fabricational device functionality. Electroforming was used to force $T_d$ to exhibit opposite polarity of switching from $B_d$. (b) State evolution of the total stack through application of voltage pulses across the aggregate stack. Exceptions to this rule are states 9 and 15, where individual switching was used for the top device, and was carried on to the total stack, in a process already described in figure \ref{fig 2}. (c,d) State evolutions of antiserially connected devices. Evidently from (c), the device state is not affected as much when using lower voltages (+2 or -2 volts), which is what was supplied to the whole stack, leading to the conclusion that higher stimulus would be needed to switch it. For the top device to switch, a pulse of increased voltage (+3 or -3 Volts) had to be separately supplied to it. This led to switching behavior seen in states 9 and 15. During states 9 to 14, whole-stack pulsing is more successful in affecting the whole stack (switching voltage +2 V) and there was a gradual decrease of $T_d$ resistance at the same time as it was increased for $B_d$. The total device evolution reflects all changes that happen to both devices. For example, transitions into states 9 and 15 are wholly dependent on $T_d$ switching, while states 1-8 and 10-14 closely follow the evolution of $B_d$.}
    \label{fig 5}
\end{figure*}

Multistate probing in this configuration does not allow full access to all possible constituent device resistive state configurations, such as when individually switching each component, with an example of this given in supplementary S2. This comes as a result of the voltage division which pushes devices to specific states. Thus each time both devices are pushed, the intermediate states are lost, and in the event of a runaway exponential resistive switching, as depicted in fig (d), the second device becomes "state-locked", as it will continue to receive a progressively diminished share of the voltage input, effectively making switching impossible.

The second case studied, and depicted in Fig. \ref{fig 5} is for when devices are connected in an antiserial manner, which could result in contrary behaviour of the two devices for a voltage pulse of any specific polarity. Figure \ref{fig 5}(a) presents the schematic for antiserially connected memristors, accompanied by an EELS image of this stack. 

In this specific scenario the stack used was the second prototype, which was fabricated with a configuration of Pt(BE)/[TiO$_x$/Al$_2$O$_3$]/Pt(ME)/TiO$_x$/Pt(TE), as is evident from the EELS image showing an Al peak in the upper Dielectric/Metal interface of the bottom device. This stack was created to take advantage of increased multistate capabilities offered by the bottom device, and to ascertain if mixing different types of memristors is possible. There were no indications that stacking devices with different active layers might have adverse effects on the switching behaviour of the entire stack. The same fundamental principles governing voltage division and switching conditions hold.

To test this case, a stack was generated, where, after electroforming both devices, one of them would increase its resistance for a specific voltage polarity, while the other would decrease it for that same polarity. The electroforming mechanism responsible for deciding the Set/Reset polarity is outside the scope of this work, nonetheless, it has been found that forming polarity is responsible up to a certain level for switching polarity as confirmed in Supplementary figure S3.

Another important detail is that ,for this experiment, the resistive values of the devices are more important than in the previous case. Devices that have similar resistivities might exhibit resistive state changes of the same magnitude, leading to apparent total-stack inactivity, while underneath, both devices are affected. On the other hand, a significant difference in resistance, or switching voltage, will lead to resistive switching in only one of two devices. Here we present the second case, while the "similar resistances and voltage requirements" scenario is presented in supplementary figure S4.

Figure \ref{fig 5}(b) presents the result of simultaneous switching in antiserially connected devices. For the initial 8 states recorded, the bottom device followed a gradual increase in resistance while the top device had no reaction to the voltage applied. It is believed that the inertness of the top device is either due to insufficient switching stimulus, a situation exacerbated with the increase in resistive state of the bottom device, or due to it already occupying its current lowest state. Following this initial resistivity increase, the Total stack was reset, by using a pulse of the opposite polarity. Then, the top device was manually switched to a higher resistive state by applying a bespoke pulse train, with opposite polarity from the one used to switch the stack. This led to a resistive switch event, seen in the transition to state 9 of subfigure (c), which also translates to a small state change in subfigure (b). Subsequently, consecutive pulse trains were once again applied to the total stack, which led to a gradual increase of the bottom device, but with a concurrent gradual decrease of resistance in the top device. The resistive change in the top device is not extensive and we believe that the stimulus that now reaches it, due to its higher resistive value, is still not high enough to incite a strong response. Nonetheless, $T_d$ behavior is counter to $B_d$. This in effect supports the notion that antiserially connected devices have opposing behaviors, with the whole stack seemingly behaving as theorized by the memristive fuse effect, although it seems more efficient to individually switch the individual devices, considering the apparently fine balance required for meeting the switching conditions of both devices at all times. Next, $B_d$ is once again reset in state 14 (Fig. \ref{fig 5}(d)), by applying a reset pulse to the total stack, and, afterwards, the top device is also set to a lower resistive state, once again, manually. Finally, another set/reset/set cycle is carried out on the entire stack. The bottom device responds to the stimulus applied, while the top device is once again unresponsive.

\section{Conclusion}

A double stack (M-I-M-I-M) memristive configuration was fabricated which combines two devices and the ability to individually switch them by using a middle electrode (ME), or the ability to simultaneously switch them by applying voltage pulses to the top (TE) and bottom (BE) electrodes of the device. Each individual device can be separately tuned, with resulting resistive changes carried over to the full stack. Full stack resistance is dependent on the nonlinearity of individual device IVs and thus is generally not equal to the sum of constituent device resistances, but rather the sum of their static resistances at the voltage shares they receive from the divider. Individually switching the devices and then registering the resistive state of the total stack can lead to increased resistive state resolution and consequently device memory density, even though the intrinsic state resolution of certain device technologies may exceed measurement resolution. Full stack switching can either switch both devices at the same time if they have similar resistivities, or only the most resistive one. Constituent device resistive state will influence the voltage balance in the voltage divider, thus deciding the amount of stimulus they will receive. 

In the case of serially connected devices of similar resistive range, switching will take place in both devices as long as the stimulus they receive is enough to change their state. For antiserially connected devices, evidence points towards a "memristive fuse" type of behavior whereby due to opposing changes in resistive states across the devices, the allocation of voltage in the divider can swing much more widely than for serial devices -in principle-. In the case of similar resistive states (and switching threshold voltages) in antiserially connected devices, the total stack may function as an attractor with unchanged resistive state, while the individual devices are impacted by the voltage stimulus applied. In either case, the existence of a second memristor in serial connection may, under the right conditions, function as a surge protection mechanism, in the same way as using a resistance in series, protecting each other from catastrophic failure. Finally, in the case of devices with the same polarity, simultaneous reset has been observed, which is a powerful tool to cut reset times in memristor arrays by half.

In conclusion, we have demonstrated in silico a 3-terminal component consisting of 2 serially connected, vertically stacked, memristors, which can be used both as a `fuse' and as individual devices. We have shown that: a) the cointegration of the two devices and b) the fact that they share an electrode does not cause any significant deviation from the theoretical behaviours that are expected when examining independent devices, either in isolation or in serial/antiserial configuration as was done in previous work. Furthermore, we show that it is possible to obtain full-stack switching in these co-integrated devices; the co-integration process does not skew the required switching conditions sufficiently to introduce catastrophic complications. We hope that this work will help the community develop such 3-terminal devices as basic components for applications that can use either their individual or their fuse-configuration properties in the future.

\section{Methods}

\subsection{Device Fabrication}

Device composition consists of two vertically aligned memristors which share one electrode (Middle Electrode - ME) of the stack. To achieve this, five distinct deposition steps were required, one for each layer of the complete device. Patterning of all steps was carried out by negative tone photolithography. A post-lithography surface clean was conducted by using O$_2$ plasma, through Reactive Ion Etching. Devices were deposited on top of a 200 nm Silicon dioxide (SiO$_2$) insulating layer, which was thermally grown on top of 6-inch silicon wafer substrates.

In steps 1-3-5 platinum electrodes were deposited by electron beam evaporation, with all electrodes having an average thickness of 12nm. For step 1 an adhesion layer consisting of a thin Titanium film (10nm) was deposited before the deposition of bottom electrodes. Following each metal deposition step a lift-off process was followed by submerging the substrates in N-Methyl-2-pyrrolidone (NMP). Thus, the three distinct electrodes were formed, Bottom Electrode (BE), Middle Electrode (ME) and Top Electrode (TE).

During steps 2 and 4 the active layers for Top and Bottom devices were deposited, for which magnetron sputtering was used. Two different material configurations were explored in this work. In case one, steps 2 and 4 had an active layer (AL) consisting of TiO$_x$, while in the second case an additional Al$_2$O$_3$ was grown on top of the TiO$_x$ in step 2 only. This enabled a first look into combinations of different types of memristive devices. In this case the TiO$_x$/Al$_2$O$_3$ active layer device was selected as it had proven to possess good device characteristics in a previous study \cite{Stathopoulos_multibit}. Deposition power used for the TiO$_x$ layer was 2 kW and for the  Al$_2$O$_3$ layer, 100 W. Flow rates were 8 sccm for O$_2$ and 35 sccm for Ar gas. A lift-off step was carried out after sputtering.

\printbibliography

\section{Acknowledgements}

The authors acknowledge the support of the EPSRC FORTE Programme Grant (EP/R024642/1), the RAEng Chair in Emerging Technologies (CiET1819/2/93), as well as the EU projects SYNCH (824162) and CHIST-ERA net SMALL.

\end{document}